# Effect of shape on void growth: a coupled Extended Finite Element Method (XFEM) and Discrete Dislocation Plasticity (DDP) study

Muhammad Usman*[1], Sana Waheed*, Aamir Mubashar*

*Department of Mechanical Engineering, School of Mechanical and Manufacturing Engineering (SMME), National University of Sciences and Technology, Islamabad 44000, Pakistan

## Abstract

Voids are one of the many material defects present at the microscopic length scale. They are primarily responsible for the formation of cracks and hence contribute to ductile fracture. Circular voids tend to deform into elliptical voids just before their coalescence to form cracks. The principle aim of this study is to investigate the effect of void shape on the micro-mechanism of void growth by using Discrete Dislocation Plasticity simulations. For voided crystals, conventional DDP produces a continuous slip step throughout the material even if a dislocation escapes from a non-convex domain. To overcome this issue, the Extended Finite Element Method (XFEM) is used here to incorporate the displacement discontinuity. Different aspect ratios of elliptical voids are considered under uniaxial and biaxial deformation boundary conditions. The results suggest that voids having the largest surface area tend to have maximum growth rate as compared to void with lower surface area, *i.e. "larger is faster"*. Under biaxial loading, a higher magnitude of strain hardening, and void growth rate are observed as compared to uniaxial loading. The results also suggest that the orientation of slip planes as well as voids, affect the overall plastic behavior of the voided-ductile material. Furthermore, circular void tends to induce minimum growth rate but have the maximum strain hardening effect as compared to other void shapes under both loading conditions. The results of this study provide a deeper understanding of ductile fracture with applications in manufacturing industry, aerospace industry and in the design of nano/micro-electromechanical devices i.e. NEMS/MEMS.

---

[1] Corresponding author: muhammadusman.pg@smme.edu.pk

## Introduction

Ductile fracture at the macroscopic level is the collective result of various interconnected phenomena occurring at the micro and nano material length scales, including void nucleation, growth and coalescence, and dislocation motion. It is important to study the interplay of theses crystal level phenomena to better understand ductile fracture. Although voided crystals have been studied using many continuum-based theories and techniques, such as J-2 flow theory [1], strain gradient plasticity theory [2] and crystal plasticity finite element method [3], these models do not naturally incorporate the dynamics of discrete dislocations and other crystal level mechanisms. Meanwhile, at the other end of the spectrum, atomistic level simulations have also been utilized to investigate the effect of dislocation nucleation and glide on void growth [4-7], but these are computationally expensive even for nano material length scales such that larger material volumes cannot be realistically studied. Discrete dislocation plasticity (DDP) [8] simulations, on the other hand, bridge the gap between atomistic and continuum length scales, allowing researchers to investigate the micromechanisms of void growth efficiently. To help matters further, experimental studies [9] have shown that most voids are present at the microscopic length scale, making DDP a suitable tool for the study of voids.

Hussein *et al.* [10] compared the response obtained from strain gradient plasticity and discrete dislocation plasticity for arrays of rectangular voids in a single crystal. However, dislocation escape from the non-convex geometry of the voided crystal was not explicitly included in this work. The modelling of dislocation escape through a non-convex domain is not straight forward and needs to be dealt with using advanced techniques which will be discussed shortly. To circumvent this issue Huang *et al.* [11] analyzed a one-fourth voided-crystal using symmetric boundary conditions, whereas Liang *et al.* [12] studied void interaction with a Mode I crack by considering the void boundary to be impenetrable to dislocations. Thus, these afore-mentioned studies use assumptions, *i.e.* symmetry and dislocation pinning at the boundary, to study the void growth in a single crystal which can lead to non-physical effects being introduced into the simulations.

The displacement field induced by dislocation escape through a void is different than the displacement field induced by dislocation escape through a convex grain boundary. Generally, in conventional discrete dislocation plasticity (DDP) methodology, dislocation escape is treated by placing the escaped dislocations at infinity along the slip plane, to obtain the equivalent displacement field of the escaped dislocation. However, as the voided crystals are a non-convex domain and the slip planes are discontinuous at the void boundary, the slip step formed after a dislocation leaves the void boundary cannot be continuous *i.e.* throughout the material. Therefore, for non-convex domains, i.e. crystals containing voids, cracks, notches, the discontinuous slip step left behind by the leaving dislocation can be incorporated by coupling the methodology either with the method of embedded discontinuities [13, 14] or the Extended Finite Element Method (XFEM) [15, 16]. As an example, the difference in deformation between conventional DDP and XFEM-DDP formulation for some non-convex domains is illustrated in Figure 1.

Romero *et al.* [13] proposed a new DDP formulation coupled with the method of embedded discontinuities to simulate the deformation of a voided crystal. Later, Segurado and Llorca [17, 18] used the proposed methodology of Romero *et al.* [13] to study the effect of void size and crystal lattice orientation on growth of voids with circular cross-section. However, in the authors' opinion, the Extended Finite Element method (X-FEM) is increasingly preferred over the method of embedded discontinuities because of the former's versatility, ease of implementation and availability in commercial software. Recently, Liang *et al.* [16]

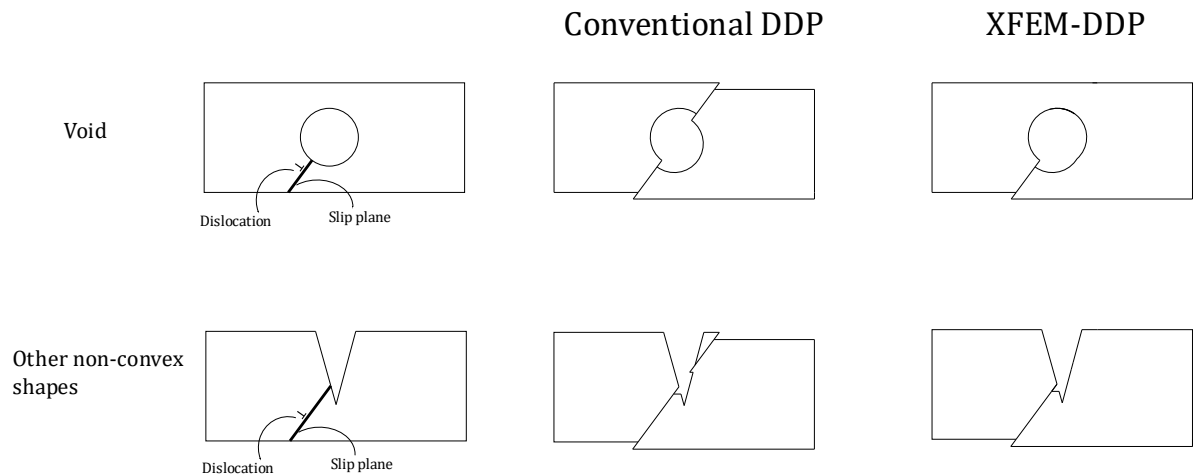

*Figure 1: Deformation in non-convex shape using conventional DDP vs XFEM-DDP formulation*

developed a new efficient XFEM-based DDD scheme, by using core-enrichment and introduction of virtual elements, which complicate the superposition scheme.

In this study, we propose a new XFEM-DDP coupled superposition scheme in which the XFEM method is used only to treat the escaped dislocations, thus removing the need for enrichment functions to capture dislocation core effects and virtual elements like Liang *et al.* [16]. The novelty that makes this new superposition scheme different from the XFEM-based DDD scheme of Liang *et al.* [16], is the dealing of dislocations inside the domain and the dislocations that has already left the domain. In the later, there is an introduction of virtual elements and the dealing of dislocations (both inside and outside the specimen) as a step discontinuity. All of the dislocations are treated as discontinuous/enriched fields containing two enrichments functions i.e., step and dislocation core enrichments, which complicates the scheme. In the present methodology, there is no core enrichment and only the escaped dislocations are dealt with Extended Finite Element Method (XFEM). We then use the proposed methodology to study our chosen non-convex domain of interest, i.e. a voided single crystal. Moreover, previous studies are restricted to only circular and square void shapes. The most important shape of the voids, just before their coalescence to form cracks, is elliptical. No previous studies have analyzed the deformation and growth rate of elliptical voids. Hence, the analysis of elliptical voids is still obscure. The second principle aim of this study is to use our developed XFEM-DDP scheme to investigate the effect of void shape on the micromechanism of void deformation and growth. For this purpose, the deformation of a single grain with an elliptical void cross-section is studied while varying its aspect ratio.

## Methodology

### *Planar Discrete Dislocation Plasticity*

The conventional DDP methodology is introduced first and then the XFEM-based coupling will be discussed later. According to conventional planar model of discrete dislocation plasticity, developed by Van der Giessen and Needleman [8], the original boundary value problem is divided into two sub-problems *i.e.* discrete dislocations (DD) based subproblem (∼) and a corrective boundary value problem (^) solved using Finite Element (FE) method. Superposition of the DD (∼) and FEM (^) subproblems is then used to calculate displacement, strain and stress fields at each increment as:

$$u = \tilde{u} + \hat{u}$$

$$\varepsilon = \tilde{\varepsilon} + \hat{\varepsilon} \tag{1}$$

$$\sigma = \tilde{\sigma} + \hat{\sigma}$$

Analytical relationships [8] are used to calculate the (~) fields of individual dislocations in an infinite medium at their current position whereas the standard Finite Element (FE) method is used to calculate the FE (^) fields which corrects for the boundary conditions of the original boundary value problem.

Sources are treated as point entities randomly distributed in the crystal. When the resolved shear stress on a particular source exceeds a critical value of $\tau_{nuc}$ within a time period of $t_{nuc}$, a dipole of dislocations with burger vectors $\pm b$ is nucleated at a nucleation distance of $L_{nuc}$. Dislocations are only allowed to glide on pre-defined slip systems. After nucleation of dislocations their motion on the slip planes is governed by a so-called Peach-Koehler force. The magnitude of the force acting on the $I$th dislocation along the slip plane is given as given as:

$$f^I = n_i^I \cdot \left( \hat{\sigma}_{ij} + \sum_{J \neq I} \tilde{\sigma}_{ij}^J \right) \cdot b_j^I \tag{2}$$

where $n_i^I$ is the unit normal vector out of the plane on which dislocation is gliding, $b_j^I$ is the burger vector of the dislocation and $\tilde{\sigma}_{ij}^J$ is the stress induced by the other $J$th dislocation on the $I$th dislocation. The dislocation glide velocity $V_{gln}^I$ is determined from the Peach-Koehler force using a linear mobility law as follows:

$$V_{gln}^I = \frac{1}{B} f^I \tag{3}$$

where coefficient of drag i.e. $B = 10^{-4}$ Pa.

In most prior DDP studies, dislocation escape through a specimen is incorporated by placing the escaped dislocations at a very far "fictitious position" along the slip plane [19]. This approach is suitable so long as the dislocation escapes from a convex domain which can be simply divided up to two regions i.e. one above the slip plane and the other below the slip plane. However, when a dislocation escapes from a non-convex domain such as specimens with voids, cracks etc., the displacement field induced by the escaping dislocation is different from the displacement field discussed earlier in the conventional DDP formulation. The grain cannot be divided into "above" and "below" regions [13]. Figure 1 shown previously illustrated the difference between conventional DDP formulation and the XFEM-DDP coupled methodology proposed in this study. The methodology is detailed in the following section.

### XFEM-DDP scheme for dislocation escape

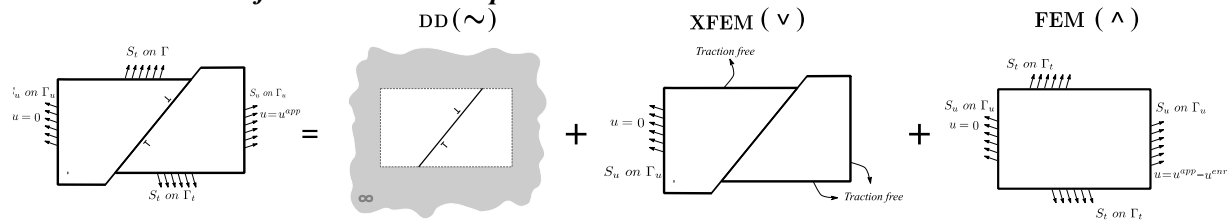

Figure 2: XFEM-DDP coupled scheme problem breakdown

A rectangular specimen undergoing uniaxial loading is used as an example to illustrate the XFEM-DDP coupling used here. The key aspect of the superposition scheme proposed in this study is to incorporate the XFEM methodology only for escaped dislocations. Instead of the conventional DDP superposition scheme outlined earlier, here the problem is divided into three sub-problems as shown in Figure 2.

The three sub-problems are: discrete dislocation DD subproblem ($\sim$) containing the infinite medium fields for dislocations remaining with the specimen, XFEM subproblem ($\vee$) dealing with the discontinuous step caused by the escaped dislocations and lastly, the standard FEM ($\wedge$) subproblem correcting for the applied boundary conditions of the original boundary value problem. Importantly, the DD subproblem deals with the dislocations inside the specimen whereas the XFEM subproblem deals with the dislocations that have escaped from the crystal, such that effects related to the analytical dislocation fields are treated within the DD subproblem and the XFEM scheme only deals with the discontinuous slip step.

Now, at every increment, the new superposition scheme, changes to:

$$u = \tilde{u} + \check{u} + \hat{u}$$
$$\varepsilon = \tilde{\varepsilon} + \check{\varepsilon} + \hat{\varepsilon} \qquad (4)$$
$$\sigma = \tilde{\sigma} + \check{\sigma} + \hat{\sigma}$$

DD ($\sim$) and FE ($\wedge$) fields contribute towards the continuous variables fields whereas the enriched XFEM field ($\vee$) incorporates the discontinuous fields. The XFEM method involves using enriched shape functions instead of the so-called standard Finite Element shape functions to incorporate materials discontinuity. These enrichment shape functions [15, 20, 21] and further details of the XFEM method are discussed below.

Consider the case of a dislocation lying on a particular slip plane that escapes through a non-convex domain as shown in Figure 3. This escaped dislocation leaves behind a slip-step along its slip plane. This slip-step creates a material discontinuity across the set of nodes in the immediate vicinity of the slip plane. The shape functions of these nodes are then *enriched* to account for this material discontinuity. More formally, enriched nodes are the nodes of elements which have at least one edge intersected by the slip plane pf the escaping dislocation. Thus, let $S_I^{enr}$ be the set of enriched nodes for the slip plane of the *I*th escaping dislocation. To demonstrate the concept of the enrichment shape function, let us consider one element that contains enriched nodes as shown in Figure 3. For the enriched element, the displacement discontinuity arising at the slip plane can be represented as:

$$\check{\boldsymbol{u}}_I(x) = \left( H(x) - \sum_{J \in S_I^{enr}} \Phi^J(x) \right) \cdot \frac{\boldsymbol{b}_I}{2} \qquad (5)$$

where $\boldsymbol{b}_I$ is the burgers vector of the escaping dislocation, $x$ is the local coordinate defined from the slip plane, i.e. positive above the slip plane and negative below the slip plane, $\Phi^J(x)$ is the linear enriched shape function for the *J*th enriched node, which is maximum (*i.e.* 1) at nodes above the slip plane and minimum (*i.e.* 0) below the slip plane and $H(x)$ is the so-called Heaviside function defined as:

$$H(x) = \begin{cases} 1, & x \geq 0 \\ 0, & x < 0 \end{cases} \qquad (6)$$

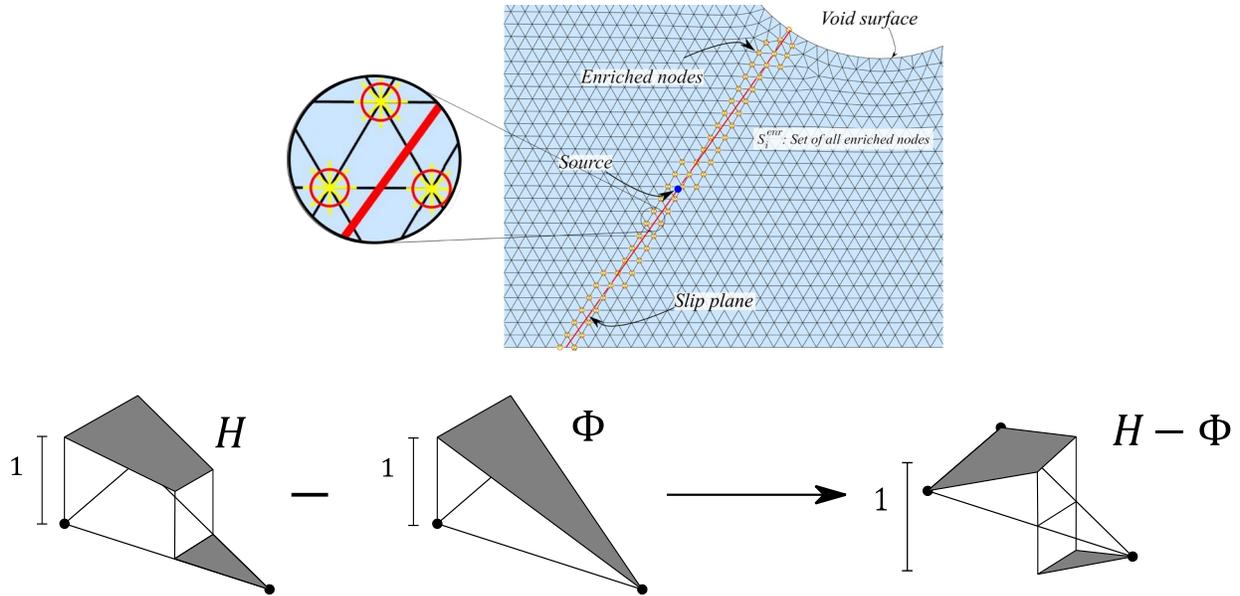

Figure 3: Enriched nodes for a single slip plane and shape function

Both the $\Phi^J(x)$ and $H(x)$ functions are shown in Figure 3 for a sample enriched element.

The discontinuous displacement field arising from the escaped dislocation leads to its respective strain field. Let the nodal displacements arising from the discontinuity be $\boldsymbol{u}^{enr}$. Then, since the gradient of Heaviside function is zero, the enriched strain is given by:

$$\check{\boldsymbol{\varepsilon}} = -\left(\boldsymbol{B}^{enr}\,\boldsymbol{u}^{enr}\right) \tag{7}$$

where $\boldsymbol{B}^{enr}$ is the enriched geometric matrix containing the derivatives of enriched shape functions. For a 3-noded triangular element (with $J$th nodes) as used in this study:

$$\boldsymbol{B}^{enr} = \begin{bmatrix} \boldsymbol{B}^{enr}_{I-1} & \boldsymbol{B}^{enr}_{I-2} & \boldsymbol{B}^{enr}_{I-3} \end{bmatrix} \quad,\quad \boldsymbol{B}^{enr}_{I-J} = \begin{pmatrix} \Phi^I_{J,x} & 0 \\ 0 & \Phi^I_{J,y} \\ \Phi^I_{J,y} & \Phi^I_{J,x} \end{pmatrix} \tag{8}$$

For the $J$th node, the nodal displacements caused by the discontinuity, $\boldsymbol{u}^{enr}$, are:

$$\boldsymbol{u}^{enr}_J = \begin{pmatrix} u^{enr}_x \\ u^{enr}_y \end{pmatrix} \tag{9}$$

The stress, $\check{\sigma}$, induced by the discontinuous strain field is given by $\check{\sigma} = C \check{\varepsilon}$, where $C$ is known as stiffness matrix. Then, for the linear triangular elements, the enriched traction ($\check{T}$) is calculated by using the enriched stress $\check{\sigma}$ simply as [22]:

$$\check{T} = \int \left(B^{std}\right)^T \check{\sigma} \, dV = A^{tri} \left(B^{std}\right)^T \check{\sigma} \tag{10}$$

where $A^{tri}$ is the area of the enriched mesh element and $B^{std}$ is the geometric matrix containing the derivatives of standard FE shape functions.

$$B^{std} = \begin{bmatrix} B^{std}_1 & B^{std}_2 & B^{std}_3 \end{bmatrix}, \quad \text{where} \quad B^{std}_J = \begin{bmatrix} B^{std}_{J,x} & 0 \\ 0 & B^{std}_{J,y} \\ B^{std}_{J,y} & B^{std}_{J,x} \end{bmatrix} \tag{11}$$

Now, the global enriched traction can be used to calculate the global enriched displacement field ($\check{u}$).

$$K\check{u} = \check{T} \tag{12}$$

Finally, having obtained the displacement field of the XFEM subproblem, the FE displacement boundary condition $\hat{u}$ is corrected by using the enriched displacement according to the superposition scheme outlined before:

$$\hat{u} = u^{app} - \check{u} - \tilde{u} \tag{13}$$

where $u^{app}$ is the applied displacement boundary condition on the original boundary value problem.

### Computational and material parameters

Material properties used here are of Aluminum with Young's modulus = 70 GPa and Poisson's ratio = 0.33. Density of randomly distributed Frank-Read sources considered here is 150 µm$^{-2}$ with a strength of 50 MPa and Gaussian standard deviation of 10 MPa whereas there are no obstacles considered in this study. Magnitude of burgers vector is 0.25 nm and nucleation time for dislocations is 10 ns. Critical annihilation distance for two oppositely signed dislocation is $6b$. The distance between slip planes is taken to be $100b$. These material parameters are consistent with the literature [11, 12, 18, 23-31]

### Problem Definition

The specimen considered in this study is shown in Figure 4. An in-plane infinite cylindrical void with an elliptical cross-section in a single crystal is analyzed under both uniaxial and biaxial displacement boundary conditions. A fixed grain orientation is considered with slip systems oriented at 0º, ±54.7º with respect to the x-axis as shown in Figure 4. The initial void volume fraction is kept constant at 10% of the grain volume. The aspect ratio of the grain is kept constant (i.e. $L = W = 2.5$ µm) whereas five different aspect ratios of the elliptical void (i.e. $l/w = 3, 2, 1, 1/2, 1/3$) are considered. Three repeats are performed for each void shape to obtain the averaged response. Initially, the sample is dislocations free and dislocation nucleation is subjected to resolved shear stress on the Frank-Read sources, as discussed earlier. The grain boundaries are made impenetrable so that the dislocations can only escape through the void surfaces.

Two different types of loadings are considered here *i.e.* uniaxial deformation and equi-biaxial deformation. The uniaxial deformation boundary condition is defined as:

$$u_x = 0 \quad and \quad T_y = 0 \quad on \ x = 0 \tag{14}$$

$$u_x = U \quad and \quad T_y = 0 \quad on \ x = L \tag{15}$$

$$T_x = T_y = 0 \quad on \ y = 0 \ and \ y = W \tag{16}$$

whereas biaxial deformation loading condition is prescribed as:

$$u_x = 0 \quad and \quad T_y = 0 \quad on \ x = 0 \tag{17}$$

$$u_y = 0 \quad and \quad T_x = 0 \quad on \ y = 0 \tag{18}$$

$$u_x = U \quad and \quad T_y = 0 \quad on \ x = L \tag{19}$$

$$u_y = U \quad and \quad T_x = 0 \quad on \ y = W \tag{20}$$

It is important here to mention that a traction-free boundary condition is assumed on the void boundary in the present study. A constant strain rate of $\dot{\varepsilon} = 2000 \ s^{-1}$ is applied to a final strain of 1% on one or both boundaries in case of uniaxial and biaxial deformation boundary condition, respectively. It should be noted that voids with different aspect ratios have different surface areas as shown in Table 1, with AR = 3 or 1/3 having the maximum surface area and AR = 1 having the least surface area. This difference of surface areas will affect our results considerably as discussed in the Results section.

*Table 1: Surface area of different void aspect ratios*

| Aspect ratios | Surface area per unit thickness / Perimeter (μm) |
| --- | --- |
| 3, 1/3 | 3.6180 |
| 2, 1/2 | 3.133 |
| 1 | 2.8025 |

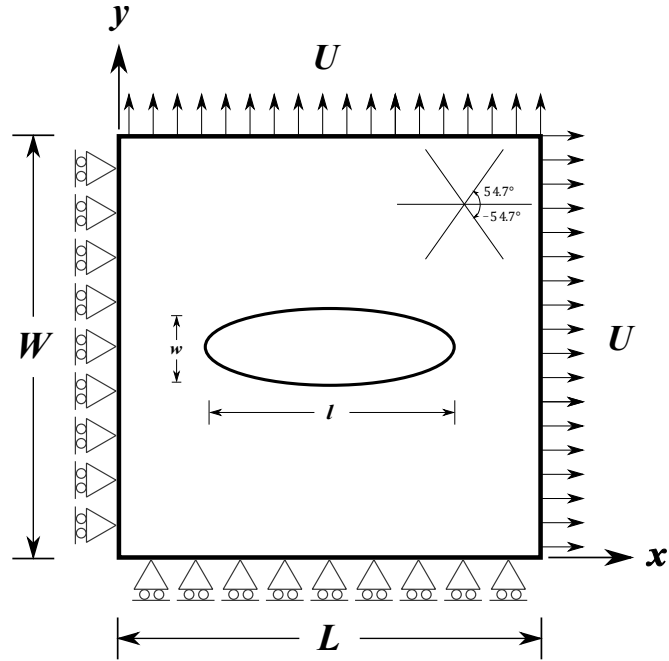

*Figure 4: Sample problem with biaxial deformation boundary condition*

## Results and Discussion

First, the developed XFEM-DDP code is validated against a conventional DDP code for a convex domain. A rectangular grain, with dimensions of 12×4 μm and a single source placed at a slip plane (oriented at 60º), is analyzed under uniaxial loading condition as shown in Figure 5. Dislocations are only allowed to escape through the single slip plane. The stress-strain results of the coupled XFEM-DDP code are compared with the conventional DDP formulation. The results from both simulations match exactly, thus validating the newly developed XFEM-DDP code. Similar exact validation is obtained for a convex domain containing random source distribution.

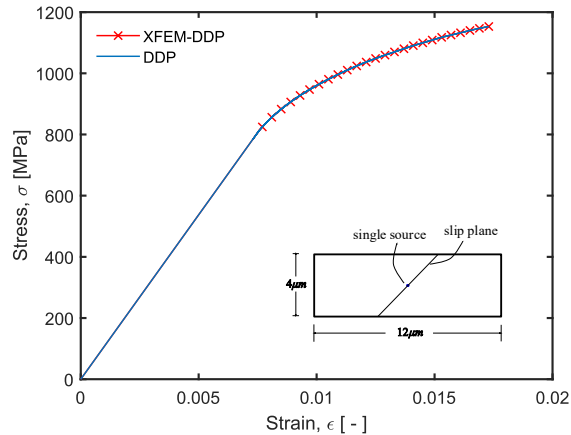

*Figure 5: Convex domain validation between conventional DDP and XFEM-DDP scheme*

Once the code has been validated, we move to the principle aim of this study which is to analyze the effect of void shape on their growth. Two different boundary conditions are considered here *i.e.* uniaxial and biaxial deformation boundary condition. Simulations are run to a total applied strain of 1%. Voids with elliptical cross-section are studied, with varying aspect ratio (AR). Deformed shapes of the voids at the end of the simulation is shown in Figure 6 a, b, c, d and e for AR = 3, 2, 1, 1/3, 1/2 respectively. As expected, larger deformation and growth is observed in case of biaxial loading as compared to uniaxial loading condition. In case of uniaxial loading, the voids tend to deform along the direction of the applied boundary condition whereas under biaxial loading condition, the original void shape is approximately retained. Considering the cross-sectional shape of the voids, from here onwards, the voids are referred to as being divided into three categories depending upon their aspect ratios: horizontal (*i.e.* AR = 3, 2), circular (*i.e.* AR = 1), and vertical (*i.e.* AR = 1/3, 1/2) voids.

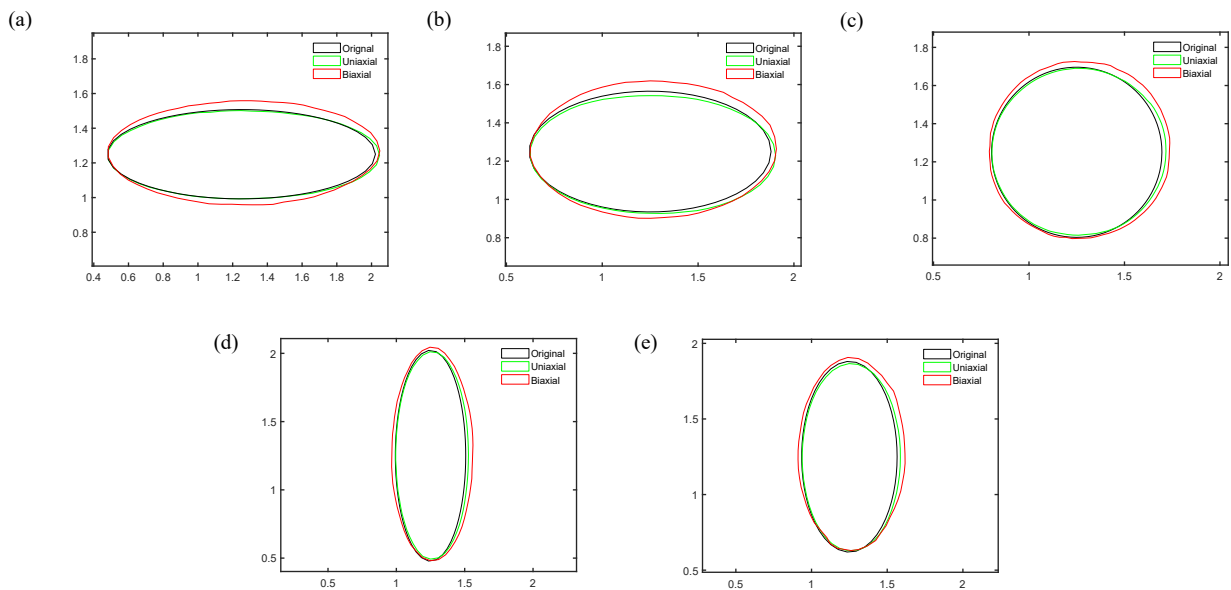

*Figure 6: Deformed void shape at the end of simulation for AR = (a) 3 (b) 2 (c) 1 (d) 1/3 (e) 1/2*

Next, Figure 8 shows the normalized void volume fraction ($f/f_o$) which is used here as a measure of void growth due to the applied loading. Under biaxial loading (Figure 8a), the voids with greater surface area correspond to a higher growth rate, with AR = 3 and 1/3 being the most deformed voids and AR = 1 being the least deformed void. In the case of uniaxial boundary condition (Figure 8b), vertical voids *i.e.* AR = 1/2 and 1/3 tend to show larger growth rate as compared to horizontal voids *i.e.* AR = 3 and 2. For vertical voids, the major axis of void is normal to the direction of applied uniaxial boundary condition thus causing the applied loading to "open up" the void. It is important here to observe that circular void *i.e.* AR = 1 tends to have minimum growth rate under both loading conditions perhaps because it provides minimum surface area for dislocations to escape.

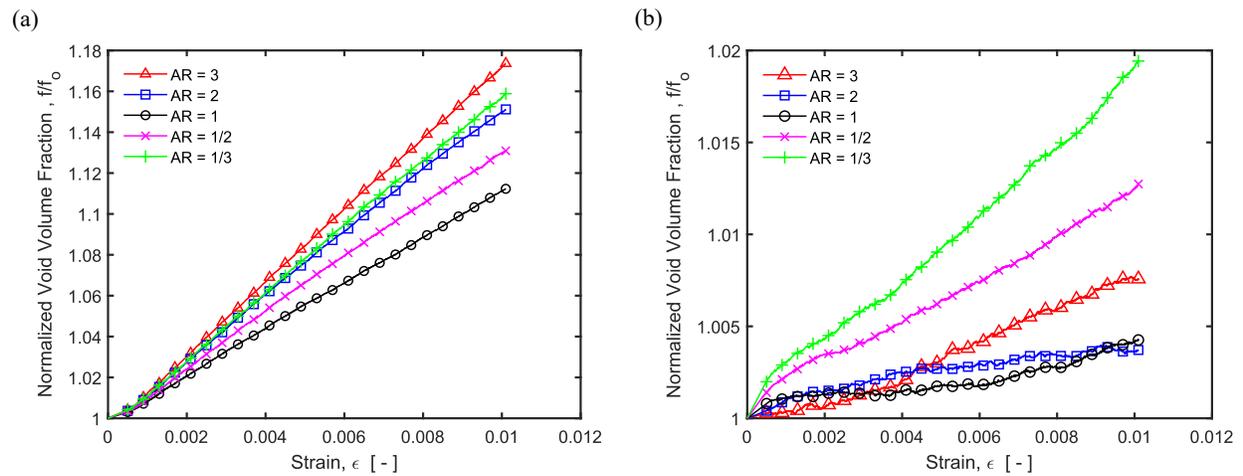

Figure 8: Normalized Void Volume Fraction for (a) biaxial and (b) uniaxial loading

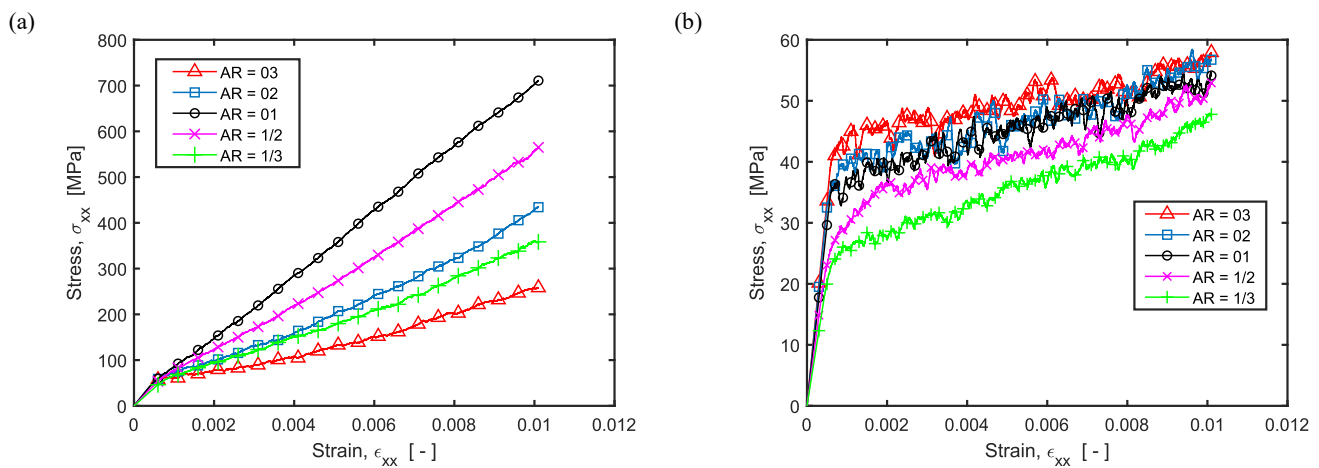

Figure 8: Normal stress vs. strain plots for (a) biaxial loading in x-direction (b) uniaxial loading in x-direction

Normal stresses ($\sigma_{xx}$), for different boundary conditions, are shown in Figure 8. For the case of biaxial loading (Figure 8a), it is observed that circular cross-section voids (*i.e.* AR = 1) have the highest stresses developed by the end of loading and the largest magnitude of strain hardening, followed by AR = 2, 1/2, and then lastly by AR = 3, 1/3. It should be noted that this trend corresponds to increasing surface area going from AR = 1 to 3 (see Table 1). Thus, for equi-biaxial loading, it can be concluded that the strain hardening effect tends to decrease as we increase the surface area of the void which is also observed by Segurado *et al.* [17]. It should also be noted here that the results for the equivalent ARs, i.e. AR = 3, 1/3 and of AR = 2, 1/2, are not the same because of plastic anisotropy introduced due to the slip systems. A similar trend is followed by $\sigma_{yy}$ (which is not shown here for the sake of brevity) as of $\sigma_{xx}$ for different aspect ratios of voids.

In the case of uniaxial loading condition, as shown in Figure 8(b), a lower magnitude of yield strength is observed for vertical voids whereas circular and horizontal voids tend to have higher magnitude of yield strength. Furthermore, it appears there is negligible strain hardening under uniaxial loading as compared to the biaxial loading condition.

Dislocation structure, under biaxial loading, at the end of simulation for different aspect ratios is shown in Figure 9. It is important to mention here that although the schmid's factor for the horizontal slip system under the applied loading is zero, but the stress fields arising from neighbouring dislocations coupled with the high stress concentration of the void, results in nucleation of dislocations on some horizontal slip planes. The phenomenon of high stress cocentration and dislocation nucleation at the top and bottom edges of voids was also observed in MD simulations by Xu *et al.* [5]. The stress state causes some sources to activate so much that there is a "dislocation avalanche" observed at the slip system oriented at 0° particularly for vertical voids (see Figure 9 (d) and (e)), which results in unrealistic results. This "dislocation avalanche", on the horizontal slip plane, was also observed in DDP simulations by [10, 12, 18] in the viscinity of voids.

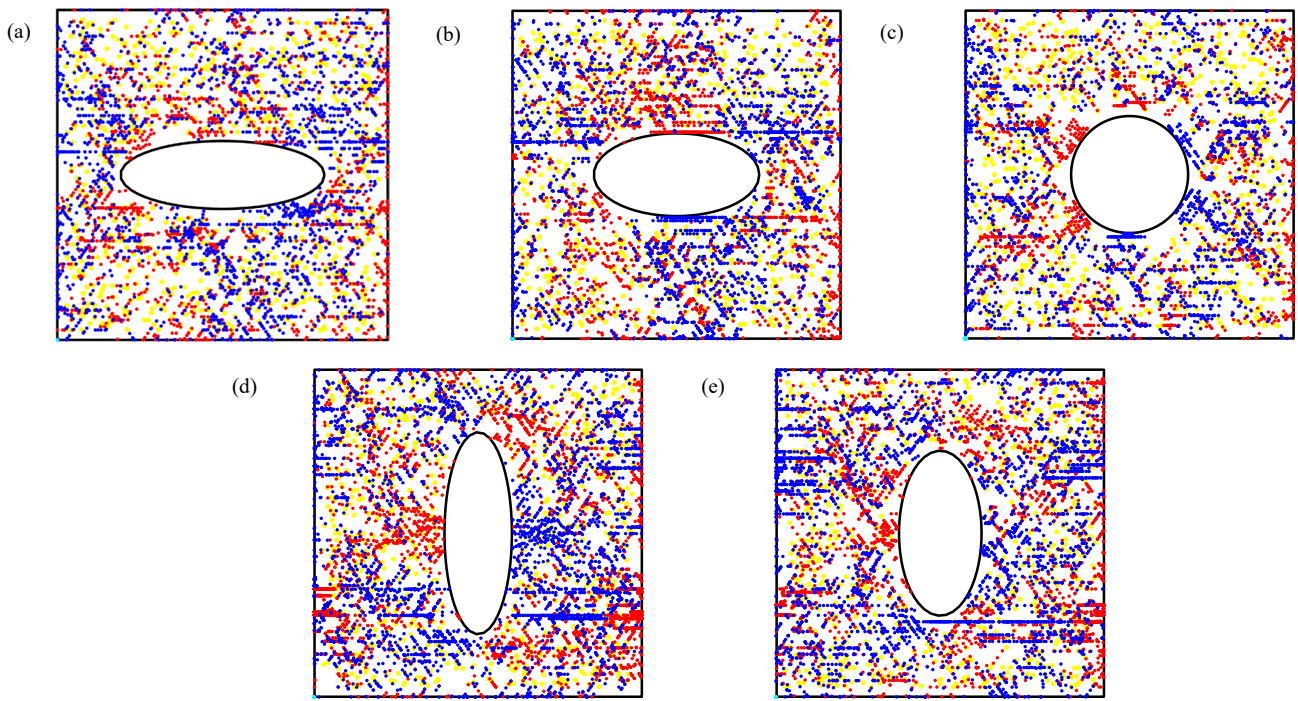

Figure 9: Dislocation structure for different aspect ratios
Sources are represented as yellow, two types of edge dislocations with oppositely signed burgers vectors are represented by red and blue colours.

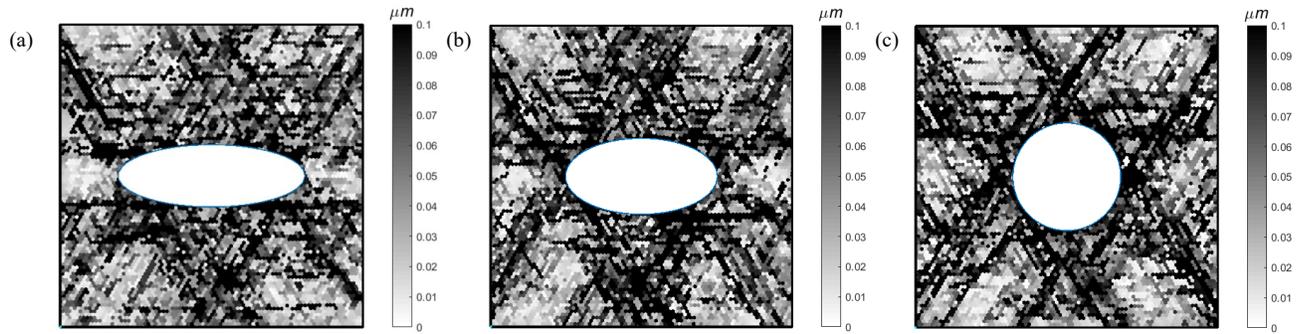

Figure 10: Slip distribution for AR = (a) 3 (b) 2 (c) 1 under biaxial loading

Under biaxial loading, the total slip distribution caused by dislocation motion inside the specimen is shown in Figure 10. It can be seen that acute and obtuse slip planes, *i.e.* ±54.7º, are more active as compared to the horizontal slip planes. Also, total slip caused by the motion of dislocations on the horizontal slip plane, due to "dislocation avalanche" at the top and bottom edges of voids, can also be seen.

Dislocation density, inside the grain, is also shown in Figure 11. An interesting fact is observed that the dislocations density is not significantly affected by the different void shapes considered in the present study, under both loading conditions.

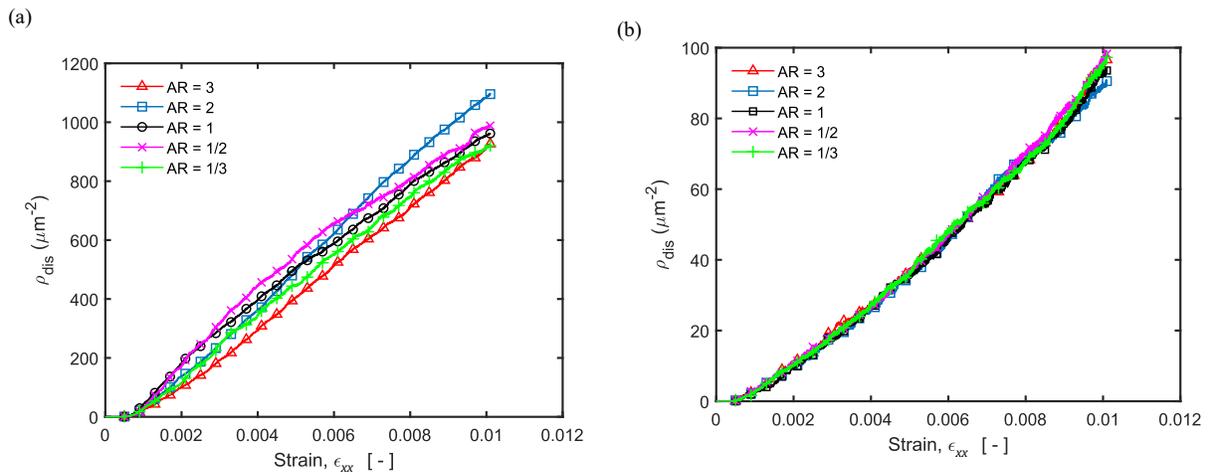

Figure 11: Dislocation density under (a) biaxial and (b) uniaxial loading

Contour plots of the von-mises stress are also plotted at the end of simulation in Figure 12 only for biaxial loading. In case of horizontal voids, stress concentration is observed in the entire region above and below the voids whereas a very symmetric stress concentration is observed around the circular void (see Figure 12c). In case of vertical voids (see Figure 12 d and e), the stress concentration is more restricted to the top and bottom edges of the void. Also, as we change the void's aspect ratio from horizontal to vertical voids, the area of "low stress" tends to increase such that, for vertical voids, a large proportion of the specimen is at near-zero von-Mises stress. In case of uniaxial loading condition (which is not shown here for the sake of brevity), stress concentration is more restricted to the top and bottom edges of the voids as also observed in MD simulations [5]. As mentioned before, the equivalent aspect ratios AR = 3 , 1/3 and AR = 2 , 1/2 have completely different results for the von-mises stress contours because of plastic anisotropy introduced due to slip systems.

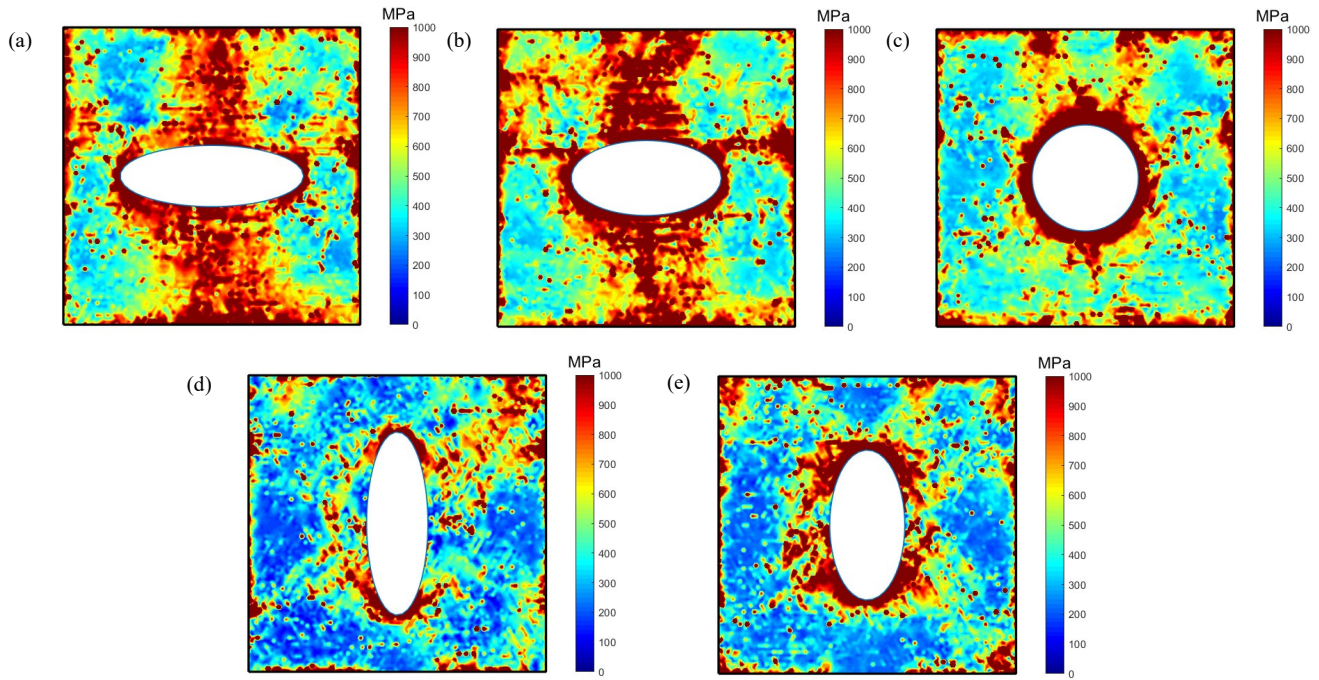

Figure 12: Von-mises stress for different aspect ratios under biaxial loading

## Conclusions

The principle objective of this study is to investigate the effect of void shape on the micromechanism of void deformation and growth by using a coupled XFEM-DDP methodology. Since the voided-grain is a non-convex domain, the dislocation escape is incorporated in the conventional DDP formulation using XFEM here. A single grain with an infinetely long cylinderical void of elliptical cross-section and different aspect ratios (*i.e.* AR = 3, 2, 1, 1/3, 1/2) is subjected to uniaxial and equi-biaxial deformation loading conditions. The aspect ratio of the grain is kept constant with an initial void volume fraction of 10% of the grain area. Following conclusions can be drawn from the present study:

- A simple and efficient methodology has been developed to incorporate dislocation escape through non-convex shapes, such as voids, cracks and notches. This methodology consists of a superposition scheme wherein the Extended Finite Element Method (XFEM) is used to tackle the discontinuous fields arising from only the escaped dislocations, whereas the conventional discrete dislocation subproblem deals with the infinite medium fields for dislocations remaining within the specimen and the standard Finite Element subproblem deals with the applied boundary conditions.

- This XFEM-DDP coupled methodology is used to analyze the effect of void shape on the deformation and growth of voids. As expected, higher stresses, strain hardening, and growth rates are observed under biaxial loading as opposed to uniaxial loading.

- It is observed that voids with an equivalent shape, e.g. an aspect ratio of 3 and 1/3, behave differently even under symmetric boundary conditions because of plastic anisotropy introduced due to slip systems. A different void growth rate and stress-strain response are observed for such equivalent AR. Therefore, the orientation of slip planes as well as voids, affect the overall plastic behaviour of the voided-ductile material.

- For a constant initial void volume fraction, the elliptical voids have a larger surface area as compared to the circular void. In addition to the effect of void and crystal orientation, it appears that with larger surface area to volume ratio, voids tend to have higher growth rate, *i.e. "larger is faster"*.

- Circular void tends to have the minimum surface area and hence the minimum growth rate under both loading conditions. But the strain hardening effect is maximum, in addition to the stresses obtained by the end of loading, in circular voids as compared to other elliptical cross-section void shapes considered in this study.

The present study is a step forward to improve the fundamental understanding of ductile fracture. This study will help to understand the micro-mechanism of voids deformation and their growth. The areas of applications include manufacturing industry, aerospace industry and nano/micro-electromechanical devices i.e. NEMS/MEMS, where the material defects at the microscopic level play a very vital role in their applications. For future research work, this study can be extended for polycrystalline materials with material heterogenities. Using the coupled XFEM-DDP methodology described here, arrays of voids, void-void interaction and void-crack interaction can also be analysed.